\documentclass[11pt,twoside]{article}
\usepackage{asp2010}

\newcommand{\gps}{\ensuremath{g_{\rm P1}}}
\newcommand{\rps}{\ensuremath{r_{\rm P1}}}
\newcommand{\ips}{\ensuremath{i_{\rm P1}}}
\newcommand{\zps}{\ensuremath{z_{\rm P1}}}
\newcommand{\yps}{\ensuremath{y_{\rm P1}}}

\begin{document}

\title{Search for and Characterization of Open Clusters Toward the Galactic 
Anti-center with Pan-STARRS1}   
\author{C. C. Lin and W. P. Chen
\affil{Institute of Astronomy, National Central University, Taiwan}}

\begin{abstract} 
We have used a star-count algorithm based the Panoramic Survey Telescope And
Rapid Response System 3$\pi$ survey data aim to identify and characterize
uncharted open clusters (OCs).  With limiting magnitudes of about 22 mag in
\gps, \rps, \ips\ bands and about 20~mag in \zps\ and \yps\ bands, our data are
100 times more sensitive than currently available surveys. We analyzed a trial
region within $20\deg\times20\deg$ field toward the Galactic anticenter and
found 1660 density enhancement regions of which 79 (out of 129) are known OCs,
and 949 are OC candidates.
\end{abstract}

\section{Introduction}
\begin{figure}
\centering 
\leavevmode
\includegraphics[width=.95\textwidth]{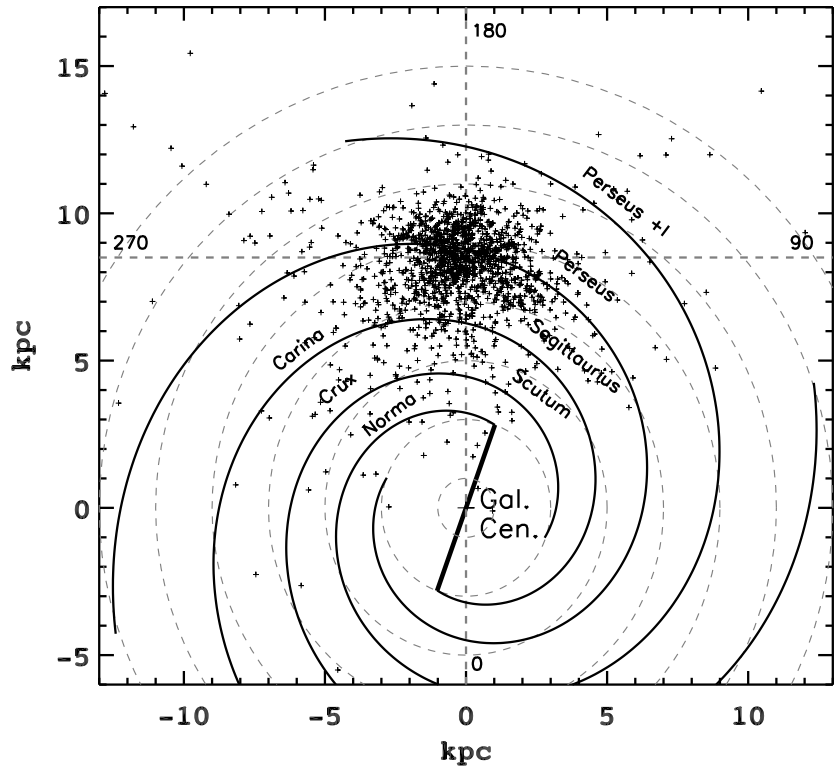}
\caption{The distribution of 1530 open clusters (OCs) in our milky way. 
The black pluses are OCs taken from \citet{dia02} updated to 2013 version.  The 
black curves were fitted to the Galactic longitude values of the tangents to 
the observed spiral arms as seen from the Sun \citep{val02}.  The distance of 
the sun is taken as 8.5 kpc (y-axis).  Dashed lines show concentric circles 
around the Galactic Center.}
\label{fig1}
\end{figure}

Hundreds of thousand open clusters (OCs) should exist currently in our Milky 
Way Galaxy, based on the number of OCs in the solar neighborhood \citep{pis06}.
However, the databases of OCs \citep{dnb01, dia02, bic03, dut03, kro06, fro07}
contain only a few thousand entries which are limited to OCs within 1~kpc (see 
Fig.~\ref{fig1}).  The discrepancy is due partly to dust extinction in the 
Galactic plane, and partly to lack of comprehensive all-sky searches for 
distant systems.\\

Star clusters are grouping of member stars in a 6-dimensional phase space in
position and motion.  Kinematic studies require special instrumentation and are
often time-consuming.  Initial identification of a star cluster via space
grouping, i.e., by the ``star-count'' technique is relatively straightforward
and has been exploited efficiently on wide-field or all-sky surveys
\citep{sch11}.  Recent work by \cite{bic03, dut03, fro07} have indeed found 
hundreds of previously unknown infrared clusters with 2MASS, some of which turn 
out to be bona fide star groups as verified by follow-up photometric studies.
A few newly found star clusters are associated with nebulosity, indicative of 
their youth \citep{lin13}.  \\

\begin{figure}
\centering 
\leavevmode
\includegraphics[width=.95\textwidth]{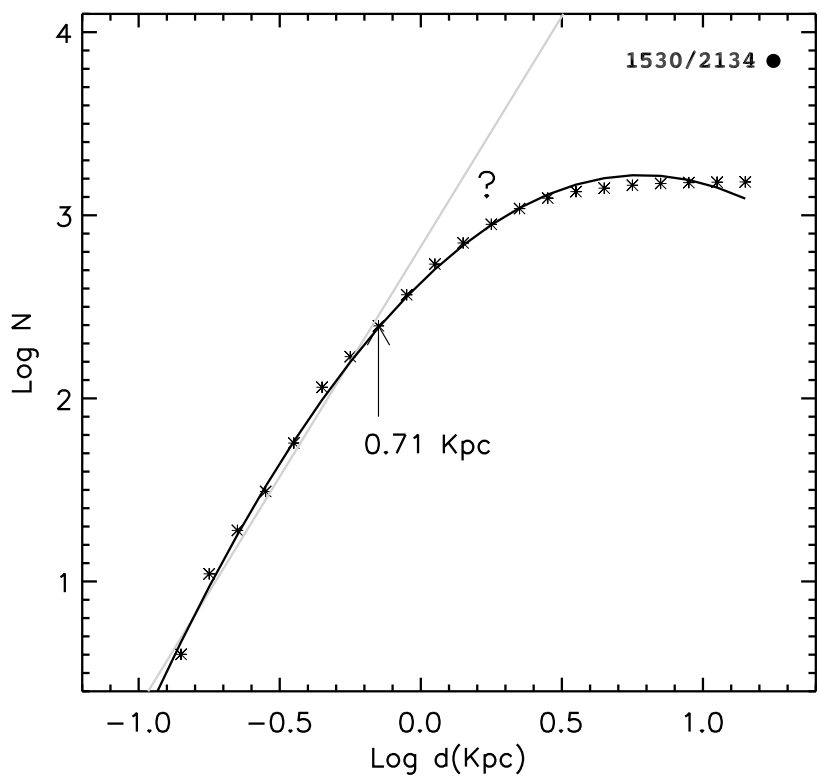}
\caption{The completeness of known OCs.  The 
asterisks represent the cumulative number of OCs with the same OC sample as in 
the left figure, at the different heliocentric distances.  The black line 
demonstrates a second order of polynomial fitting, while the grey line shows 
the expected cumulative number of OCs if uniform distribution of OCs density is 
assumed.  The completeness is at 0.71~kpc shown as the black arrow marks.}
\label{fig2}
\end{figure}

Nonetheless, a systematic analysis of star clusters from deeper optical 
photometry survey has not yet been studied.  Fig.~\ref{fig2} demonstrates the 
incompleteness properties of OCs, only 1530 out of 2134 OCs have distance 
information \citep{dia02} and the completeness limit is only at 0.71~kpc.  
Since all these OCs are analyzed with the different dataset, one needs a 
more uniform and deeper dataset to resolve the properties of further OCs.  The 
Pan-STARRS1\footnote{http://ps1sc.org/} -- Panoramic Survey Telescope And Rapid 
Response System -- $3\pi$ survey provides us a deeper and wider coverage than 
the recent sky survey, we therefore can not only discover more OCs to expand 
their distance of the completeness limit, but also estimate the properties of 
OCs more systematically and comprehensively.\\

As a training set for PS1 data, we have developed and tested a 
star-count algorithm to recognize star density enhancements toward the Galactic 
Anti-center with 400 square degrees in the PS1 $3\pi$ catalog.  Here we report 
on the completeness and limitation of our searching algorithm and preliminary 
characterization from PS1 data.

\section{Data and Analysis} 
\subsection{Pan-STARRS1 Data}
The Pan-STARRS1 (PS1) is an innovative design for a wide-field imaging facility 
developed at the University of Hawaii's Institute for Astronomy and the 1.8~m 
telescope is located on Haleakal\={a} on the island of Maui, conducting a 
multi-wavelength, multi-epoch, optical imaging survey \citep{kai02}.  The 
$3\pi$ survey covers about 30,000 square degrees and the limiting magnitudes, 
on the AB system, for each filter are 23.4, 22.8, 22.2, 21.6, and 20.1 mag in 
\gps, \rps, \ips, \zps, and \yps, respectively \citep{ton12}.  \\

We extract data within a field of view centered on ($\ell,b) = (180.0,0.0)$ 
with  $20\deg\times20\deg$ box size on PS1 $3\pi$ survey data from PS1 Desktop 
Virtual Observatory (DVO) database \citep{mag06}.  According to the 
photometric quality flags set, we exclude the extend sources and require 
objects to be detected at least five times to remove potential false 
detections.  Furthermore, we require the scattering of object's positions less 
than 200 mili-arcsec.  Finally, we have 203,226 sources out of 211,668 entries.

\subsection{Identification of Star Density Enhancements}
\begin{figure}
\plottwo{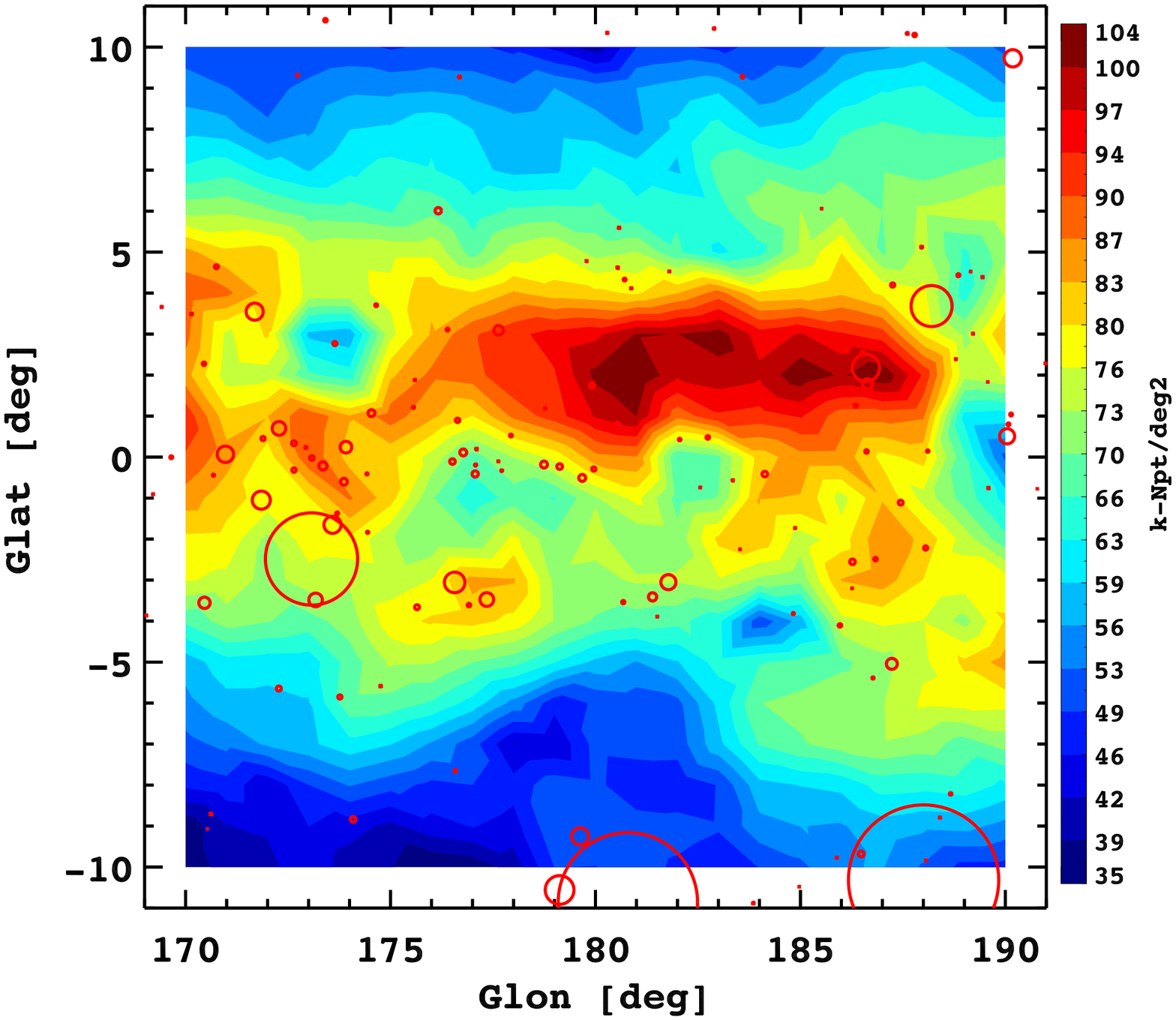}{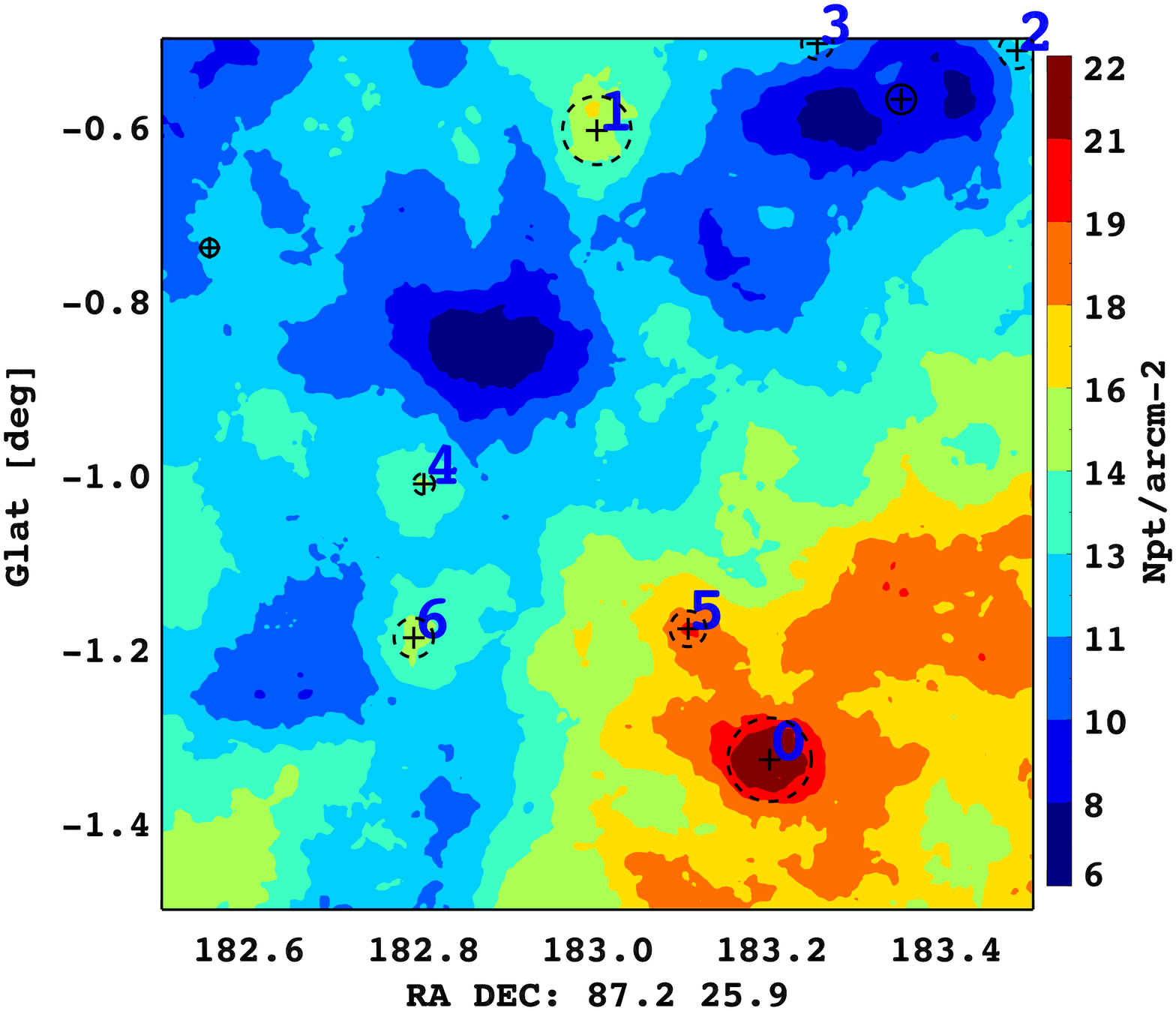}
\caption{(Left) Galactic anticenter star density maps of PS1 $3\pi$ survey 
data. A total of 129 known OCs with their published diameters are marked with 
red circles  \citep{dia02}.  (Right) One of our results shows the one square 
degree star density map at $(\ell, b)=(183.0\deg, -1.0\deg)$.  Black dashed 
circles with numbers are detected OCs with the effective diameters, and black 
pluses mark their center 
coordinates.}
\label{fig3}
\end{figure}

We choose good quality data to make the star density map whose pixel values 
are determined by counting the number of stars in a 3 arcmin radius with 10 
arcsec offset (see Fig.~\ref{fig3} left panel).  The star density maps were 
then divided into one square degrees and searched automatically for local 
density enhancements by using the SExtractor software \citep{bna96}.  \\

We found 1660 density enhancements toward Galactic anticenter in a region 
within $20\deg\times20\deg$ field.  In Fig.~\ref{fig3} right panel, we 
demonstrated a sample of our results.  The black circles are our candidates, 
there may be some contaminations from (e.g., bright stars, chip gaps, clouds 
etc.).  Therefore, a cluster candidate will be identified only if it satisfies 
with (a) the local star density is above the $5\sigma$ level of the background, 
(b) an effective radius of more than one arcmin, (c) having a positive core and 
tidal radius by fitting with the King's model \citep{kin66}, and (d) having 
more than 10 members.  Following those criterias, we then have 949 probable OC 
candidates.

\subsection{Characterization of Star Cluster Candidates}

\begin{figure}
\centering 
\leavevmode
\includegraphics[width=.99\textwidth]{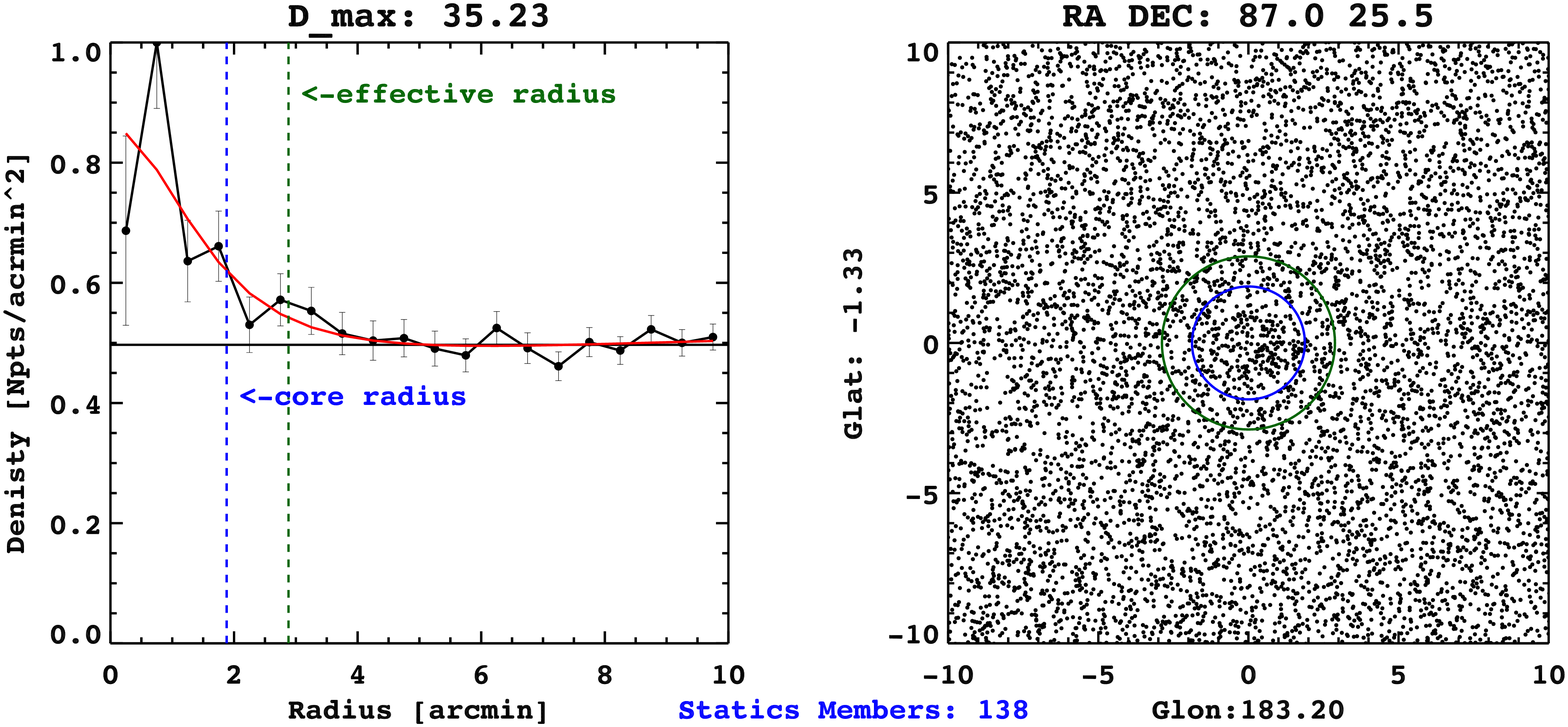}\\
\includegraphics[width=.95\textwidth]{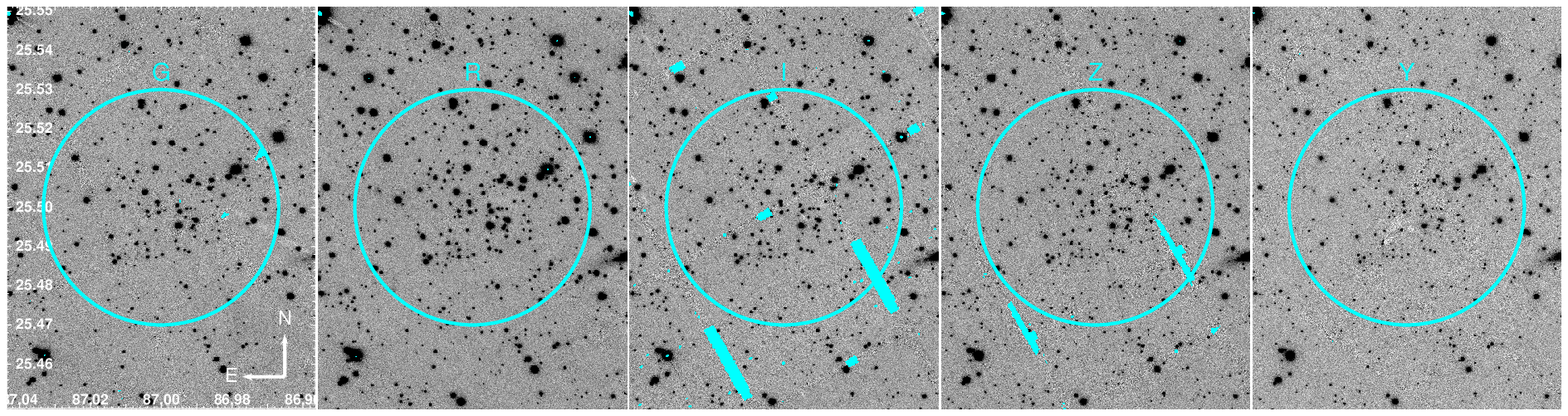}%
\caption{(Upper Left) The radial density profile. Horizontal black line 
represents the mean background star density.  Green lines demonstrates the 
effective radius, and blue lines are core radius estimated by fitting King’s 
model.  Statistic number of members are labeled.  (Upper Right) The 
identification map.  The solid circle are the same labels as in the upper left. 
(Bottom) The PS1 stacked images from left to right are \gps, \rps, \ips, \zps, 
and \yps-band images, respectively}
\label{fig4}
\end{figure}

Fig.~\ref{fig4} shows our analysis of our candidates No. 0 in Fig.~\ref{fig2} 
right panel. The effective radius is determined by the density enhancement 
region, however we considered to use the radius derived by King's model.  The 
radial density profile suggested a field density of about 17 stars per square 
arcmins, and a density enhancement within about 2 arcmin radius by fitting 
with the King's model. The central density in the cluster region reaches about 
35 stars per square arcmins, and there are a total of 138 member stars within 
the 2 arcmin cluster radius.  One can see the PS1 stacked images in the bottom 
of Fig.~\ref{fig3}, and the central region indeed shows the stars 
concentration.  The age, distance, and reddening of our candidates are not yet 
determined here, however we are improving our programs recently \citep{lin13, 
wan13}.

\section{Results and Summary}
We identified and characterized uncharted OCs based on PS1 $3\pi$ survey data.  
A total of 1660 detections were identified by our pipeline.  In our search 
field, there are 129 known OCs, for which 79 were rediscovered.  Those not 
found by our pipeline are either too extended ($> 30$ arcmin) or with 
uncertain position (offset $>3$ arcmin).  A total of 949 OC candidates have 
been identified.  The remaining 632 are unlikely real OCs because they are with 
neither enough members nor significant radial profiles.  We expect to produce a 
comprehensive OC sample out to about 3~kpc.  With PS1 photometry and proper 
motion, very low-mass members down to 0.5~M$_{\sun}$ at 3~kpc, 0.15~M$_{\sun}$ 
at 1~kpc can be identified. \\

\acknowledgements 
We acknowledge the financial support from the National Science Council grant 
NSC~102-2119-M-008-001 and NSC~102-2119-M-008-002.  The Pan-STARRS1 Surveys 
(PS1) have been made possible through contributions by the Institute for 
Astronomy, the University of Hawaii, the Pan-STARRS Project Office, 
the Max-Planck Society and its participating institutes, the Max Planck 
Institute for Astronomy, Heidelberg and the Max Planck Institute for 
Extraterrestrial Physics, Garching, The Johns Hopkins University, Durham 
University, the University of Edinburgh, the Queen's University Belfast, 
the Harvard-Smithsonian Center for Astrophysics, the Las Cumbres Observatory 
Global Telescope Network Incorporated, the National Central University of 
Taiwan, the Space Telescope Science Institute, and the National Aeronautics and 
Space Administration under Grant No. NNX08AR22G issued through the Planetary 
Science Division of the NASA Science Mission Directorate, the National Science 
Foundation Grant No. AST-1238877, and the University of Maryland.


\end{document}